\documentstyle[11pt,newpasp,twoside,epsf]{article}
\def\d#1{\displaystyle{#1}}

\def\edcomment#1{\iffalse\marginpar{\raggedright\sl#1\/}\else\relax\fi}
\reversemarginpar

\begin{document}
\title{Compton Scattering, Pair Annihilation and Pair Production
in a Plasma}
\author{Vinod Krishan}
\affil{Indian Institute of Astrophysics,
       Bangalore 560034, India}

\begin{abstract}
The square of the four momentum of a photon in vacuum is zero.
However, in an unmagnetized plasma it is equal to the square of
the plasma frequency.  Further, the electron-photon coupling
vertex is modified in a plasma to include the effect of the
plasma medium.  I calculate the cross sections of the three
processes - the Compton scattering, electron-positron pair
annihilation and production in a plasma.  At high plasma
densities, the cross sections are found to change significantly.
Such high plasma densities exist in several astrophysical sources.
\end{abstract}

\section{Introduction}

The processes of the Compton scattering, electron-positron pair
annihilation and production play an important role in the
functioning of highly energetic astrophysical sources such as
pulsars, supernovae, $\gamma$-ray bursts and the early universe
(Wolfgang, Fabian \& Giovannelli 1990; Rose 1973). The cross 
sections of these processes are used, almost
always, either in the classical Thomson limit or at best
including the Klein-Nishina corrections for high energy photons.
To the best of my knowledge, the effect of the plasma medium on
the cross sections
has not been investigated.  In this paper, I show that the
cross sections change significantly when the presence of the
plasma medium is taken into account.  The plasma medium affects
in two ways.  First the square of the four momentum of a photon
is no longer zero, it is equal to the square of the plasma
frequency, and second the electron-photon vertex now depends
upon the dielectric constant $\epsilon$ of the plasma medium.
The cross sections become functions of the plasma density n.

The dispersion relation of photons in an unmagnetized plasma is
given by $\omega^2 = \omega^2_p + \vec k ^2 c^2$, where $\omega$
is the photon frequency, $\vec k$ is the photon wavevector and
the plasma frequency $\omega_p =(4 \pi ne^2/m)^{1/2}$ in an
electron-proton plasma and $= (8\pi ne^2/m)^{1/2} $ in an
electron-positron plasma of density n.  Thus the square $k^2$ of
the photon four momentum $k = (\omega, \vec k c)$ is found to be
$\omega^2 - k^2 c^2 = \omega^2_p$ (Krishan 1999).  
The total energy associated with a wave in a dielectric (Harris
1975; Landau \& Lifshitz 1960), is modified by the
Von Laue factor $\d{\frac{1} {2\omega} \frac{\partial} {\partial
\omega} (\omega^2 \epsilon (\omega))}$ so that the vector
potential $\vec A (x)$ is defined as:

\begin{equation}
\vec A (x) = \left[\frac{2 \pi \hbar c^2} {V\left[\d{\frac{1} {2}
\frac{\partial} {\partial \omega} \omega^2 \epsilon}\right]_{\omega_{k}}}
\right]^{1/2}  \Lambda_{\vec k \sigma} \left[a_{\vec k \sigma} e^{i k.x} 
+ a^+_{\vec k\sigma}  e^{-i k.x} \right]
\end{equation}
where $a$'s are  the photon creation and annihilation operators.
For an unmagnetized plasma, the dielectric function $\epsilon = 1
- \omega^2_p/\omega^2$.  The vector potential $\vec A (\vec x)$
is modified only to the extent that the photon dispersion
relation in vacuum is replaced by the photon dispersion
relation in a plasma.  With these
changes, one can proceed to calculate the cross sections for the
three processes related by the crossing symmetry using the standard
techniques of the quantum electrodynamics (Reinhardt 1992, 1994).

\begin{figure}[h]
\plotone{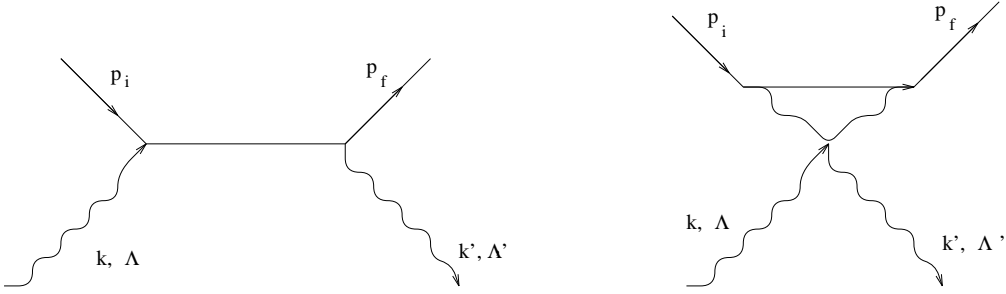}
\caption{The direct and exchange diagrams for the Compton scattering:
$e^{-}+\gamma\rightarrow\gamma'+e^{-}.$}
\end{figure}

\section{Compton Scattering}

The Compton scattering consists of the absorption of an incoming
photon with four-momentum k and polarization vector $\Lambda$ by
an electron with the emission of a second photon with
four-momentum $k^\prime$ and polarization vector
$\Lambda^\prime$.  The Feynman diagrams for this process are
shown in Figure (1).
Here, $p_i$ and $p_f$ are the four-momenta of the initial and
the final electrons respectively.

\begin{figure}
%\vskip -1.0 cm
\plottwo{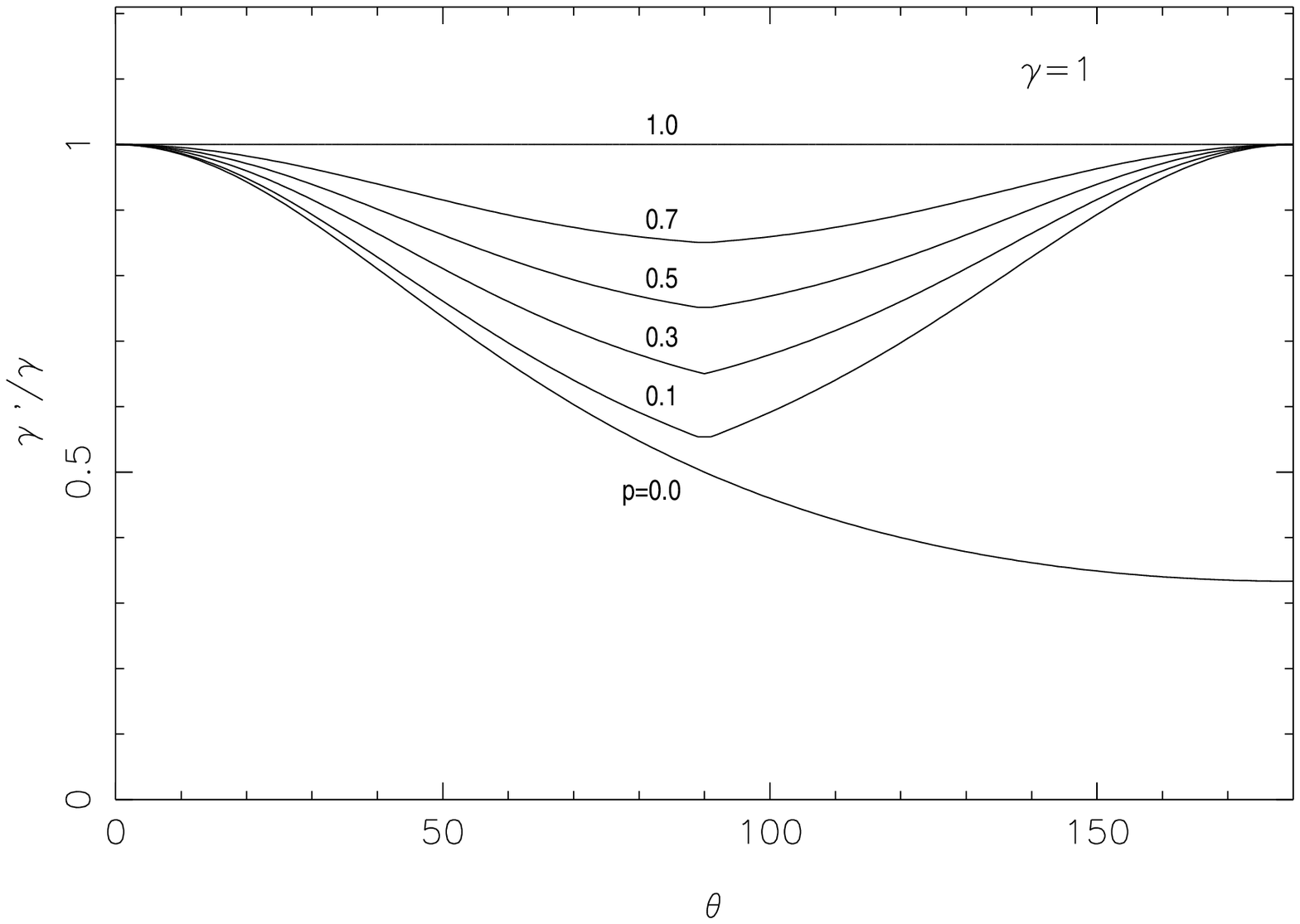}{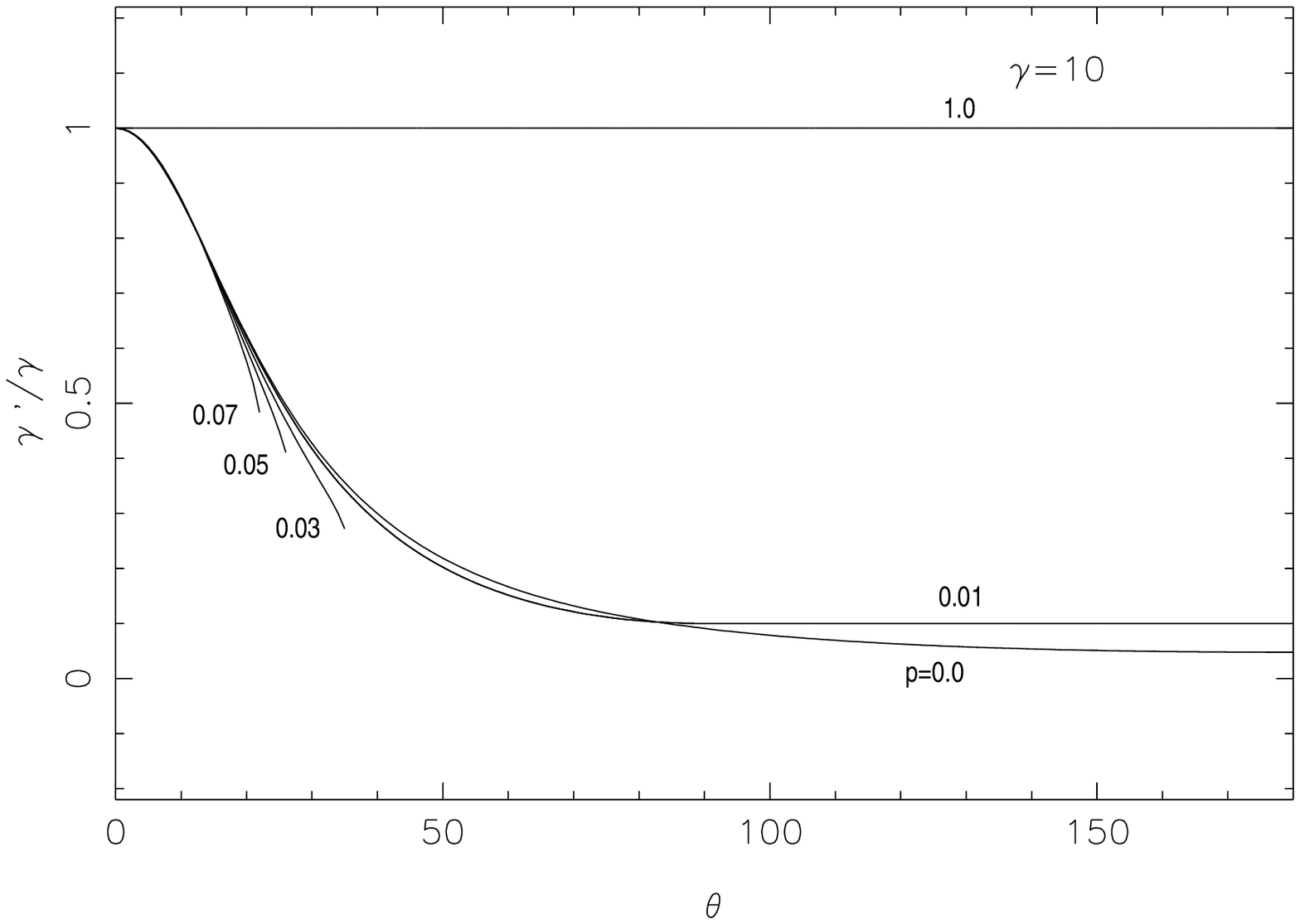}
%\vskip -1. cm
\caption{Variation of the first root $(\gamma^\prime/\gamma)$
of equation (3) vs the scattering angle $\theta$ for different
values of the plasma density parameter p and initial photon
energy $\gamma$.}
\end{figure}
\begin{figure}
%\vskip -1.0 cm
\plottwo{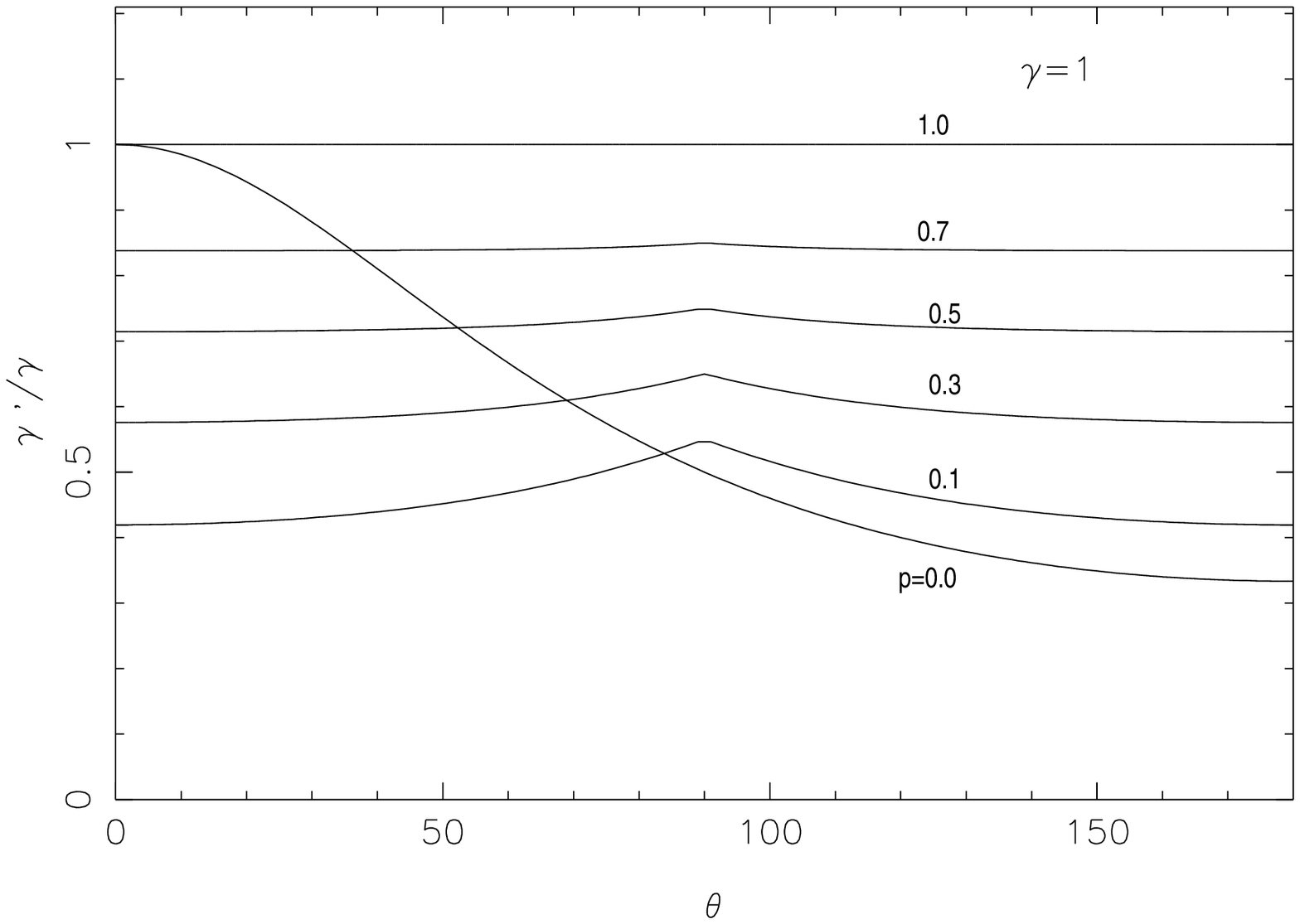}{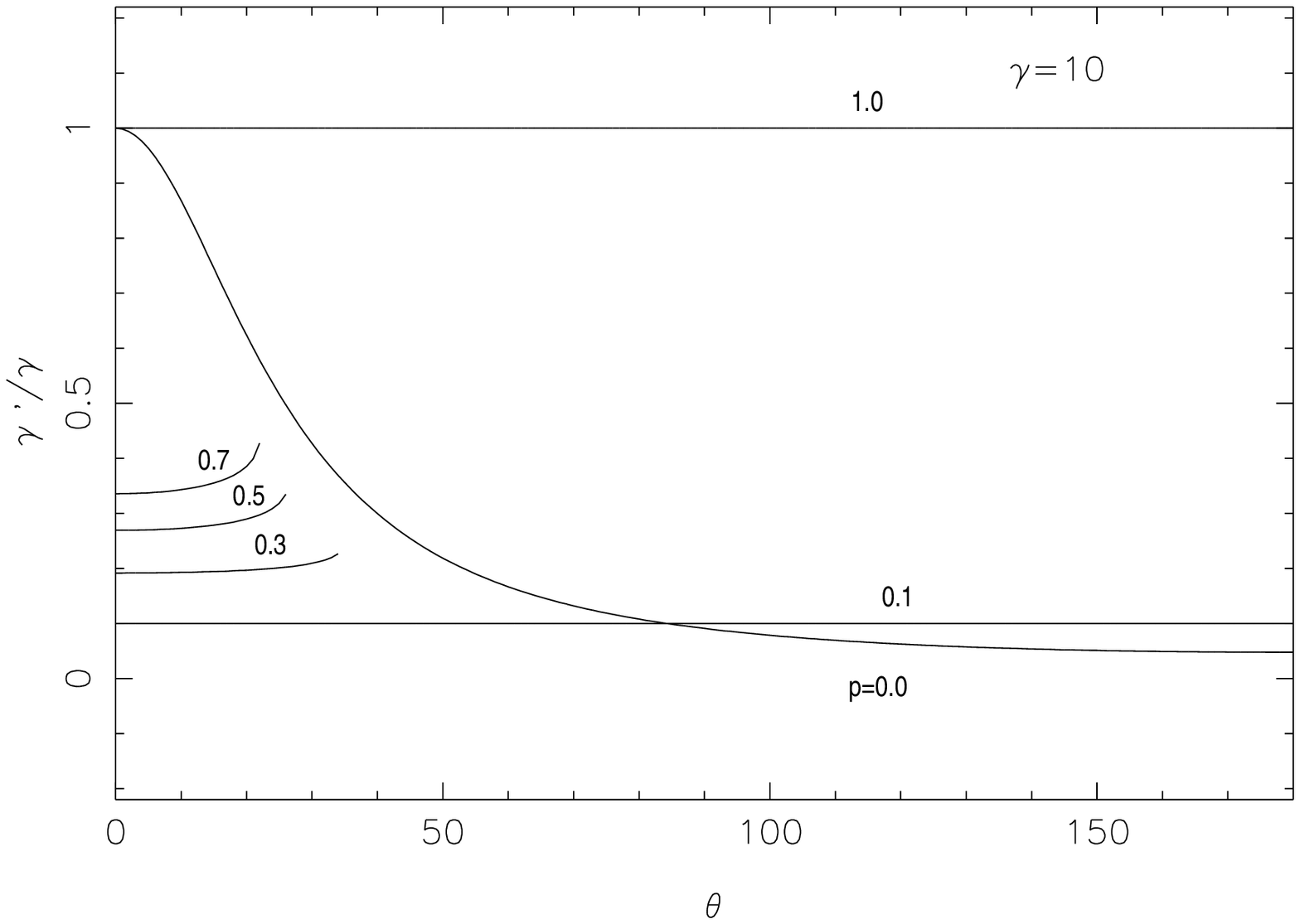}
%\vskip -1. cm
\caption{Variation of the second root $(\gamma^\prime/\gamma)$
of equation (3) vs the scattering angle $\theta$ for different
values of the plasma density parameter p and initial photon
energy $\gamma$.}
\end{figure}
\begin{figure}
%\vskip -1.0 cm
\plottwo{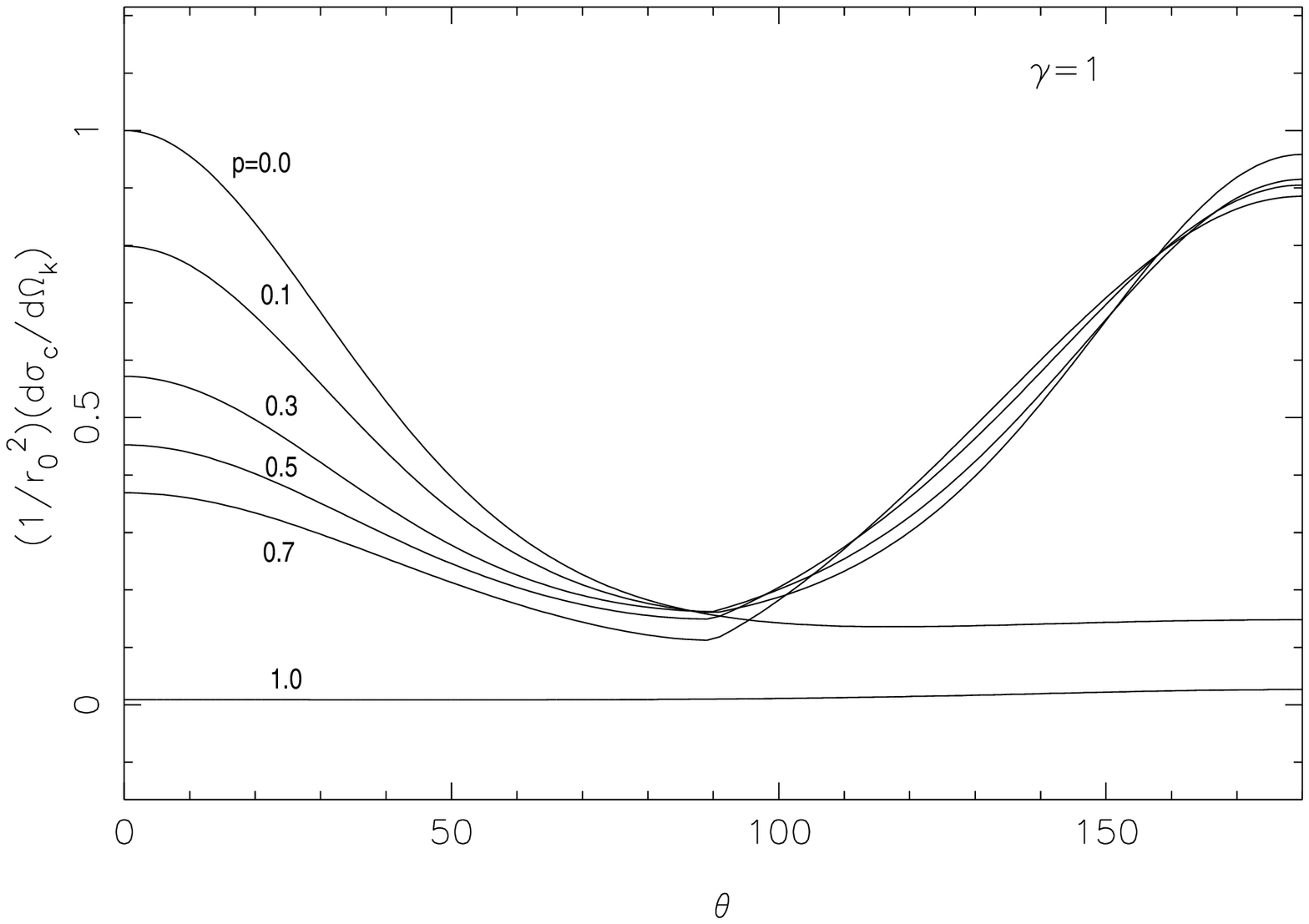}{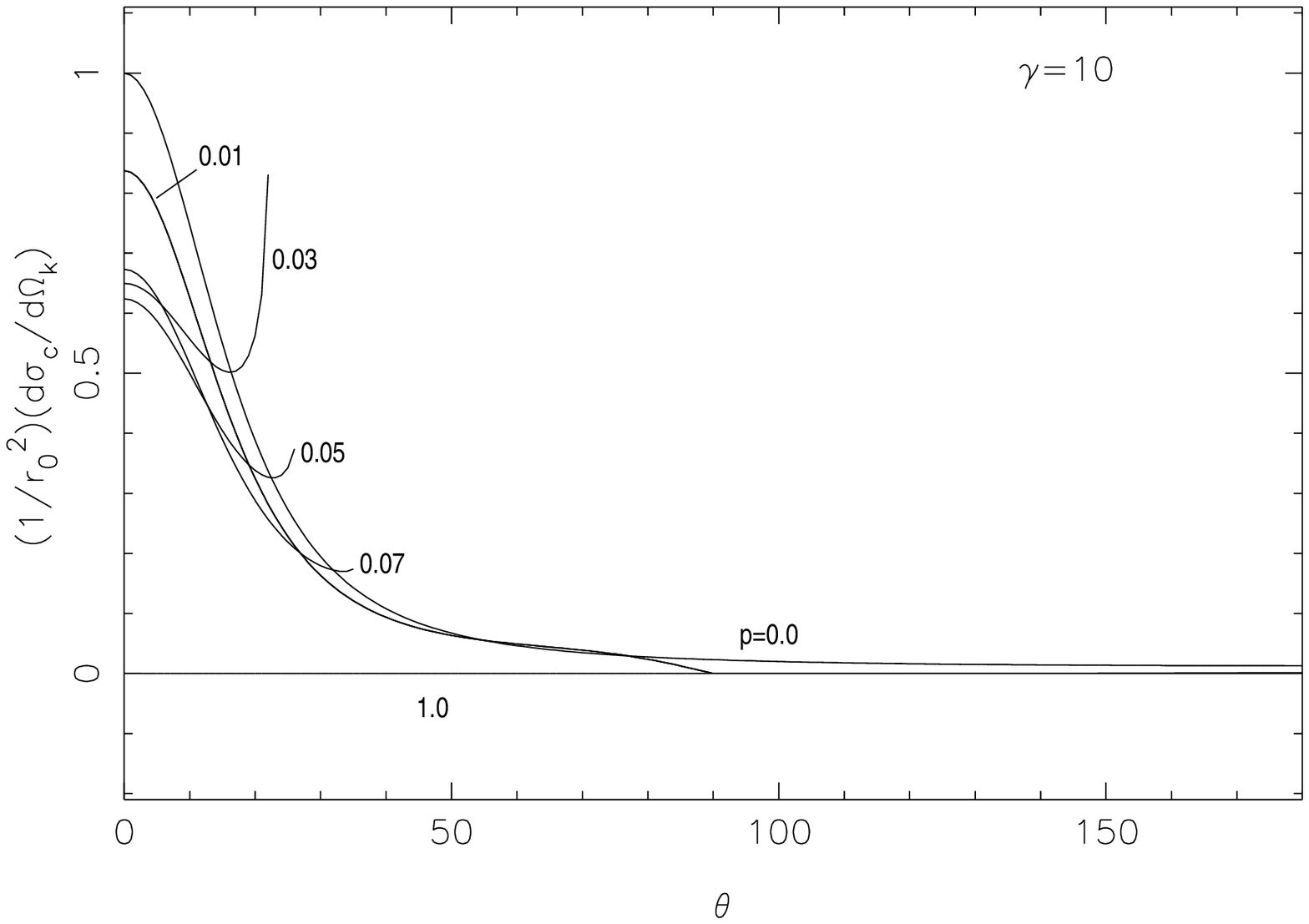}
%\vskip -1.0 cm
\caption{Variation of the differential cross section for the
Compton scattering with the scattering angle $\theta$ for
different values of the plasma density parameter p, the initial
photon energy $\gamma$ and the final photon energy
$\gamma^\prime$ corresponding to the first root of equation (3).}
\end{figure}
\begin{figure}
%\vskip -1.0 cm
\plottwo{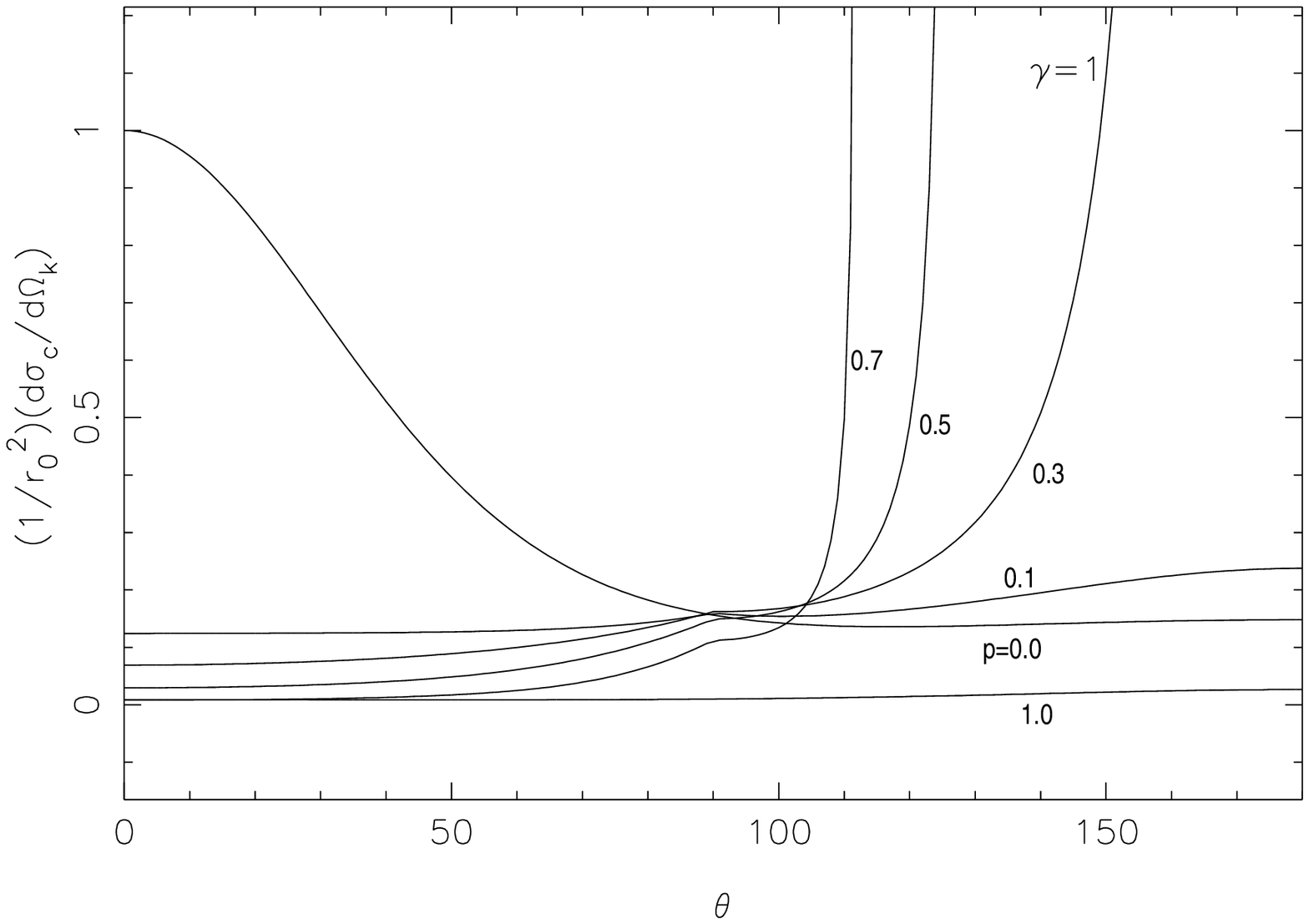}{fig5b.ps}
%\vskip -1.0 cm
\caption{Variation of the differential cross section for the
Compton scattering with the scattering angle $\theta$ for
different values of the plasma density parameter p, the initial
photon energy $\gamma=1$ and the final photon energy
$\gamma^\prime$ corresponding to the second root of equation (3).}
\end{figure}

The differential cross section for the Compton scattering
averaged over photon polarizations is found to be:
\begin{eqnarray}
&\d{\frac{d \bar \sigma_c} {d\Omega_k} \frac{1} {r_0^2}} =\left[\d{ \frac{\gamma^{\prime
^{2}}}
{\gamma^2} \frac{(1+ \cos^2 \theta)} {2\sqrt{AB}}} + \d{\frac{\gamma^{\prime^{2}}} {4
\gamma^2}} \left[ \frac{\gamma^\prime} {\gamma A} + \frac{\gamma}
{\gamma^\prime B} - \frac{2} {\sqrt{AB}} \right.\quad\quad\quad\right. \nonumber\\
&\left.\left.+ \left(1 - \d{\frac{\sqrt{A}} {\sqrt{B}}}\right)
\d{\frac{\gamma} {A}} (1-p) \sin^2 \theta 
 -\left(1- \d{\frac{\sqrt{B}} {\sqrt{A}}}\right) \d{\frac{\gamma^\prime} {B}} (1 - p^\prime) 
\sin^2
\theta \right.\right.\nonumber\\
&\left.\left. - \d{\frac{p\gamma} {AB}} \left\{\d{\frac{2\gamma} {\gamma^\prime}}\sqrt{AB} + A
\left(\d{\frac{\gamma^2} {\gamma^{\prime^{2}}}} + \d{\frac{\gamma} {\gamma^\prime}}
\right) - B \left(\frac{\gamma^\prime} {\gamma} + 1\right) -2\sqrt{AB} \right\} \right]\right] G
\end{eqnarray}
where $\gamma = \hbar \omega/mc^2$ and $\gamma^\prime = \hbar
\omega^\prime/mc^2$  are the initial and
the final photon energies related by the energy-momentum
conservation law,

\begin{equation}
  p + \frac{1} {\gamma} = \gamma^\prime/\gamma^2 +
(\gamma^\prime/\gamma) \{1-\sqrt{1-p} \sqrt{1-p^\prime} \cos \theta\},
\end{equation}
and 
\begin{displaymath}
G=\left[1+p\gamma+p^\prime\gamma^\prime\sqrt{1-p}\cos\theta/\sqrt{1-p^\prime}\right]^{-1}\sqrt{1-p^\prime}
\end{displaymath}
Here $\theta$ is the scattering angle,
$A = (1+p\gamma/2)^2, B = (1 - p^\prime
\gamma^\prime/2)^2,$ $p = \omega^2_p/\omega^2,$ $\ p^\prime =
\omega^2_p/\omega^{\prime^{2}} \ \ \hbox{and} \ \
r_0^2 = (e^2/mc^2)^2.$
%########
One can easily identify the Thomson, the Klein-Nishina and the
additional terms which depend upon the plasma frequency.
Unlike in vacuum, the initial and the final photon frequencies
are related through a quadratic expression.  For every incoming
photon, there are two possible scattered photons.  The two
values of the ratio $(\gamma^\prime/\gamma)$ are plotted
against the scattering angle $\theta$ for various values of p
and $\gamma$ in figures (2) and (3).  The first root (for plus sign)
corresponds to an increase of the frequency of the scattered
photon as the plasma density or the parameter $p$ increases 
for $\gamma=1$. For $\gamma=10$, the variations with $p$ are very small.
The second root  also increases with $p$ but lies below the $p=0$ line
for small values of scattering angle $\theta$. 
 Further the quadratic (3) has real roots only for a
limited range of scattering angle for a given value of $p$.
The variation of the two differential cross sections
corresponding to the two roots (Equation 2) with
$\theta$ for different values of the parameter $p$ and initial
photon energy $\gamma$ is shown in
figures (4) and (5).  The cross section corresponding to the
first root is found to decrease significantly
with an increase in $p$ for $\theta<90^\circ$ and increase for 
$\theta\geq
90^\circ$. For $\gamma=10$, the cross section begins to increase for
$\theta << 90^\circ$. 
Further the scattering takes place only
for limited values of the scattering angle. The cross section for the
second root (Figure 5) shows a similar behaviour but becomes very very
large for large scattering angles. This root may not represent the
physical situation.

\section{Pair Annihilation}

The annihilation of an electron with four-momentum $p_-$ and spin $S_-$
and a positron with four-momentum $p_+$ and spin $S_+$ into two
photons with four-momenta and polarizations $k_1, \Lambda_1$ and
$k_2, \Lambda_2$ is depicted by the Feynman diagrams in figure (6).
\begin{figure}
\plotone{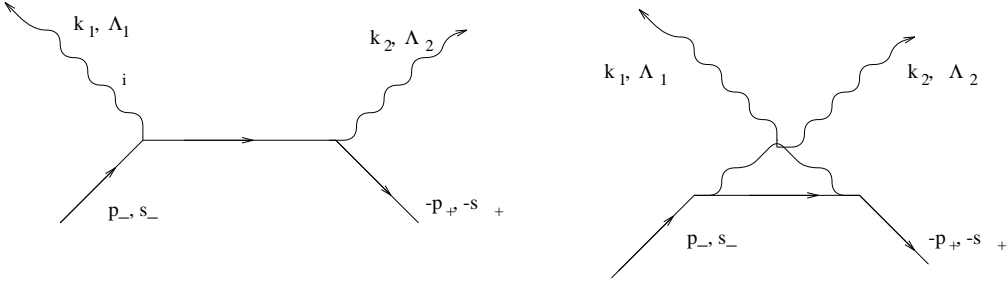}
\caption{Direct and exchange diagrams for electron-positron
annihilation : $e^- + e^+ \rightarrow \gamma_1 + \gamma_2$}
\end{figure}
\begin{figure}
\plottwo{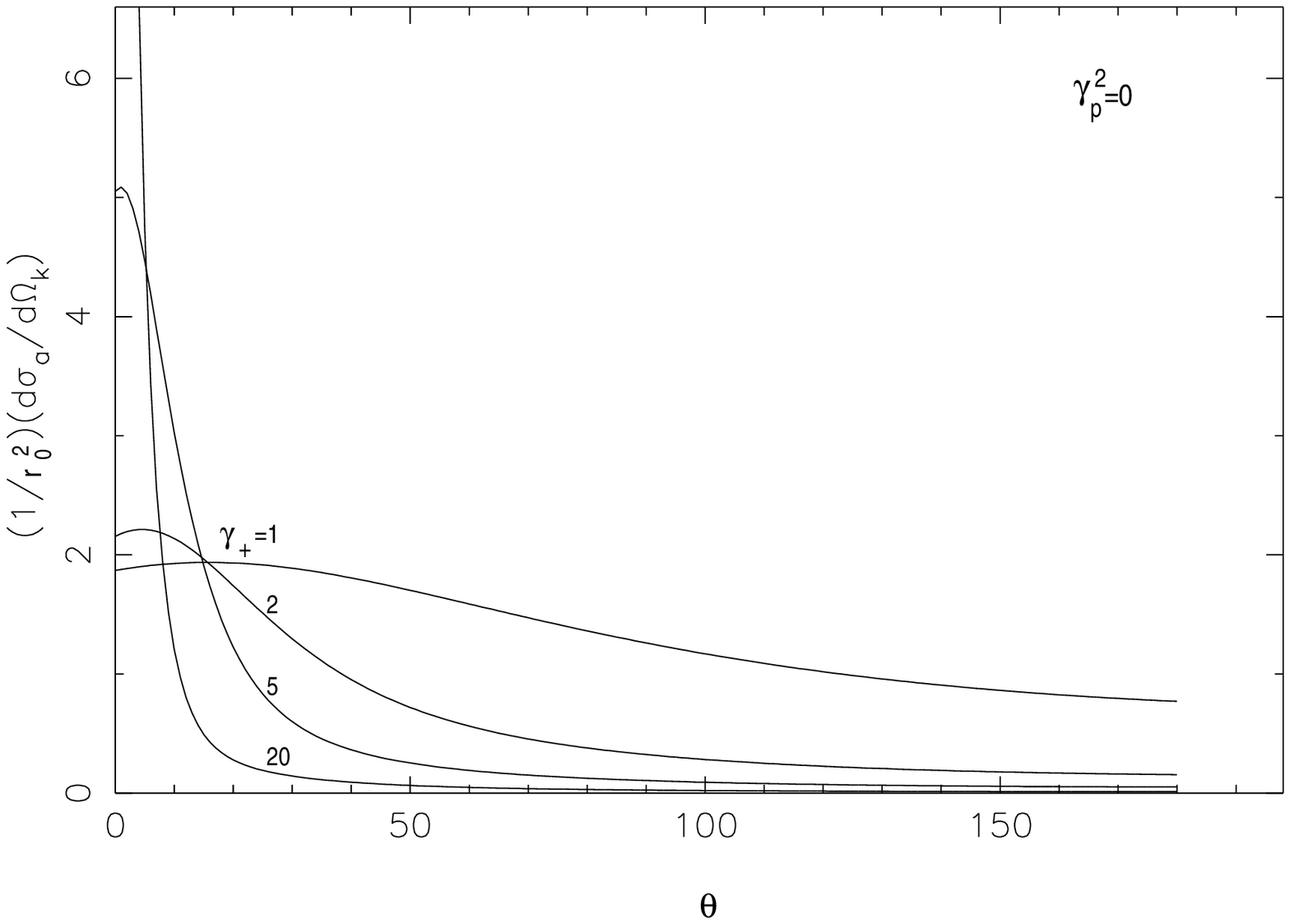}{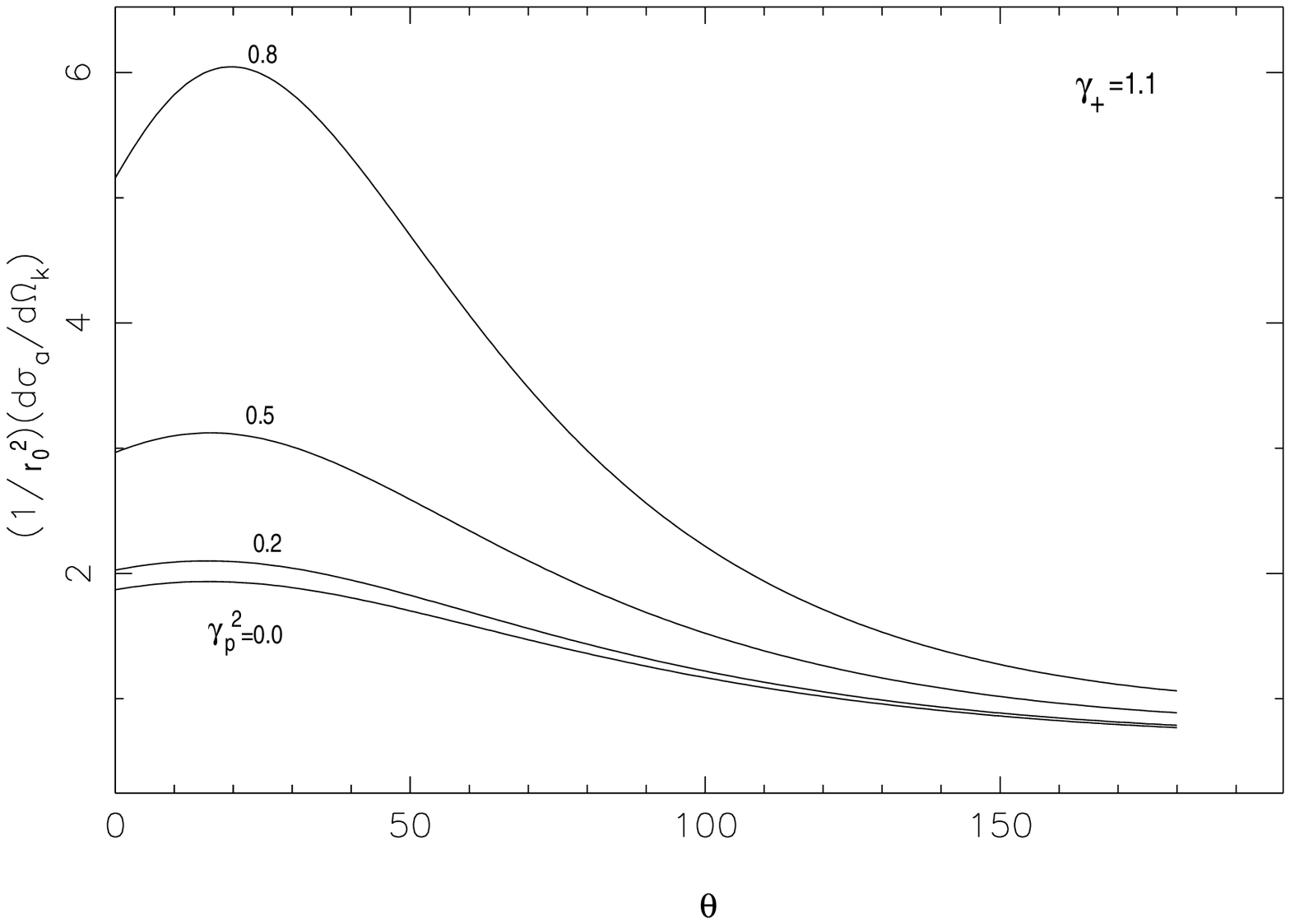}
\caption{Variation of the differential cross section for
pair annihilation with the angle $\theta$ between the
positron beam and the photon ($\gamma_1$) for different positron
energies $\gamma_+$ and plasma density parameter $\gamma^2_p$.}
\end{figure}
\begin{figure}
\plottwo{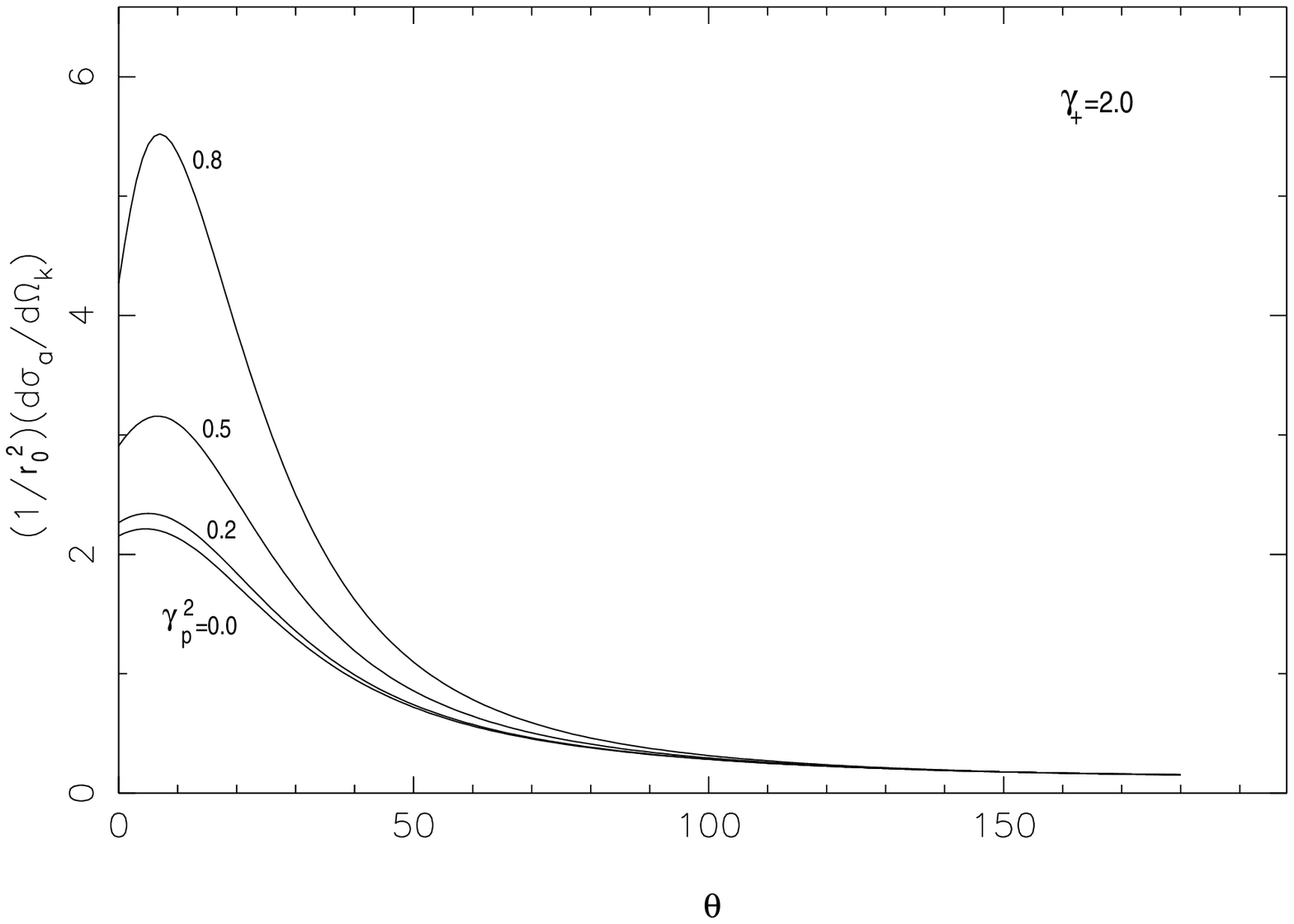}{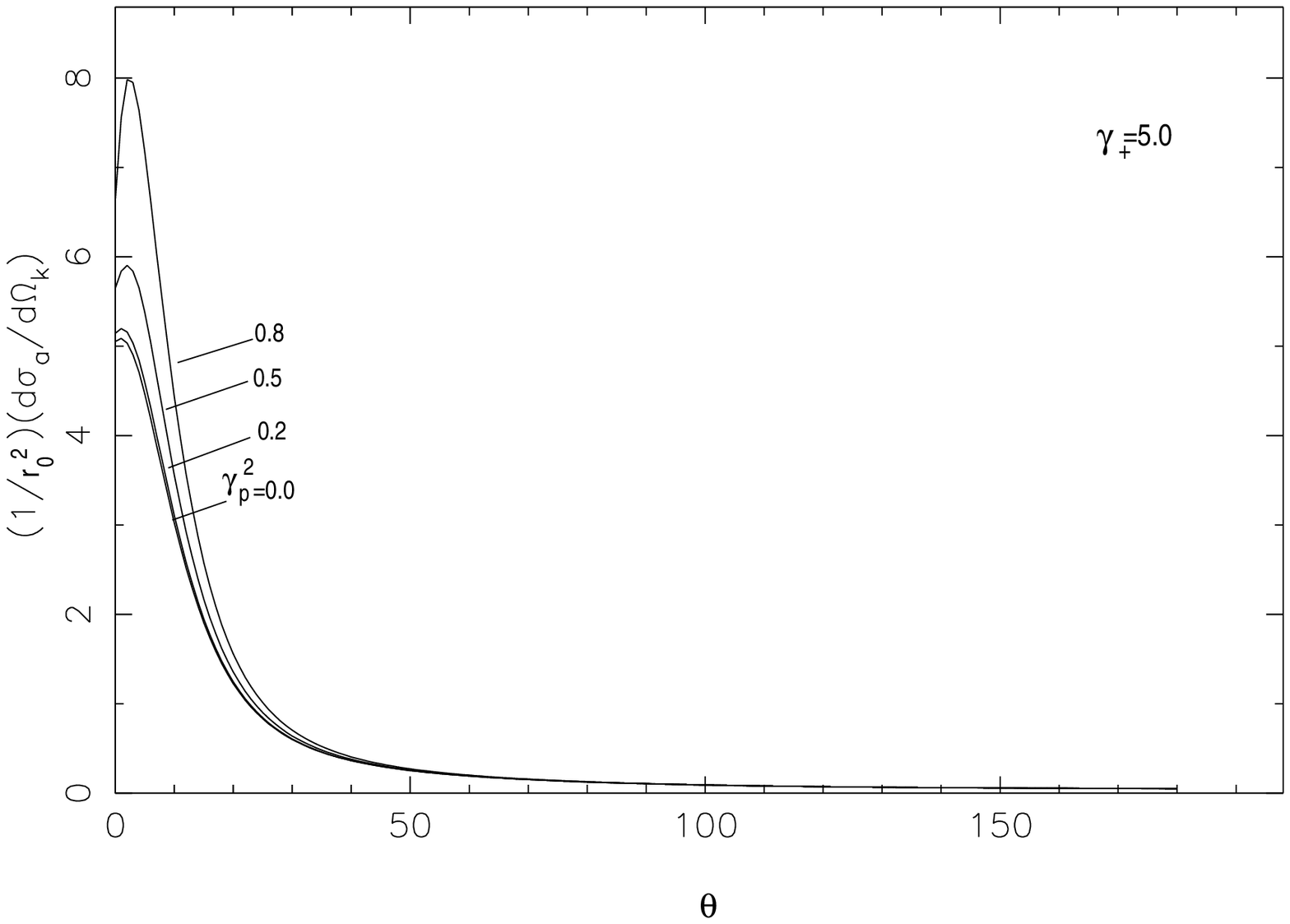}
\caption{Variation of the differential cross section for
pair annihilation with the angle $\theta$ between the
positron beam and the photon ($\gamma_1$) for different positron
energies $\gamma_+$ and plasma density parameter $\gamma^2_p$.}
\end{figure}
The similarity with the Compton scattering is obvious.  By using
the crossing symmetry (Reinhardt 1992, 1994) i.e. by replacing $p_i \rightarrow
p_-, -p_f \rightarrow p_+, -k \rightarrow k_1$ and $k^\prime
\rightarrow k_2$, in the Compton scattering matrix, one can
obtain the matrix elements for the pair annihilation process.
By appropriately taking care of the phase space factors and
averaging over the initial and summing over the final states, we
find the differential cross section for pair annihilation in the
rest frame of the electron to be:
\begin{equation}
\frac{d \bar \sigma_a} {d \Omega_k}  \frac{1} {r^2_0} =
\frac{(1 + \gamma_+) \left[\d{\frac{\gamma_2} {\gamma_1}} \d{\frac{1}
{A_1}} + \d{\frac {\gamma_1} {\gamma_2}} \d{\frac{1} {B_2}} + D \sin^2
\theta_o + E\right]\sqrt{1-p_1}}
{2 \beta_ + \{1 + \gamma_+ - (1 - p_1)^{-1/2} \beta_+ \cos
\theta\} \{1 + \gamma_+ -(1-p_1)^{1/2} \beta_+ \cos \theta\}}
\end{equation}
where $\gamma _+ mc^2$ is the positron energy, $\gamma_2 $ and
$\gamma_1$ are the energies of the two photons,
$A_1 = (1-p_1
\gamma_1/2)^2, B_2 = (1 - p_2 \gamma_2/2)^2,$
\begin{equation} D = 1/A_1 B_2
-(1-\sqrt{A_1/B_2}) \gamma_1 (1-p_1)/2A_1 - (1-\sqrt{B_2/A_1}) \gamma_2 (1-p_2)/2B_2
\end{equation}
\begin{eqnarray}
&E= (- p_1 \gamma_1/2A_1 B_2) \left\{ \left(- 2\gamma_1/\gamma_2\right)
 \sqrt{A_1 B_2} +
A_1(\gamma^2_1/\gamma^2_2 - \gamma_1/\gamma_2)\right. \nonumber\\
&\left. - B_2 (1-\gamma_2/\gamma_1)-2\sqrt{A_1B_2}\right\}
\end{eqnarray}
\begin{equation}
\sin^2 \theta_o = \frac{\beta _+ \sin \theta} {\gamma_2
(1-p_2)^{1/2}} , \ \  p_1 = \gamma^2_p/\gamma^2_1, \ \ p_2 =
\gamma^2_p/\gamma^2_2, \ \ \gamma_p = \hbar \omega_p/mc^2, \beta^2_+ =
\gamma^2_+ - 1,
\end{equation}
\begin{equation}
\gamma_2 = \gamma_+ + 1 - \gamma_1, \ \ \gamma_1 = \frac{1
+\gamma_+} {1 + \gamma_+ - (\gamma_+ - t) \sqrt{1-p_1}},\ \ 
\cos \theta = \frac{\gamma_+ - t} {\beta_+}
\end{equation}

\begin{figure}
\plotone{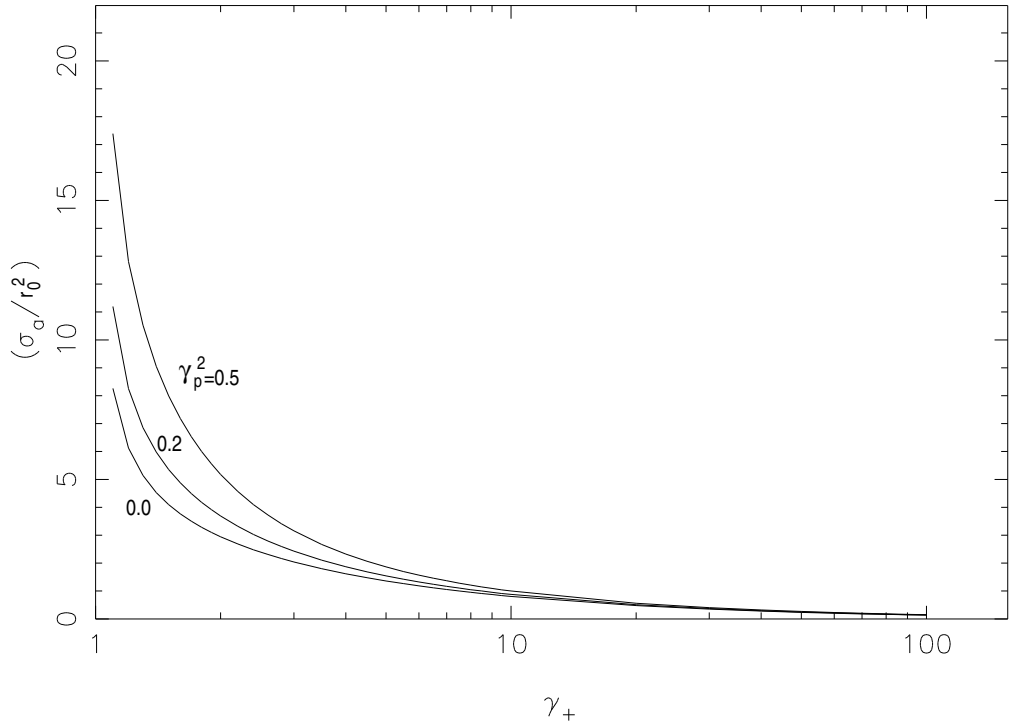}
\caption{Variation of the total cross section $\sigma_a$ for
pair annihilation with the positron energy $\gamma_+$ for
different values of the plasma density parameters $\gamma^2_p$.}
\end{figure}

The variation of the differential cross section for pair
annihilation with $\theta$, the angle between the positron beam
and the photon ($\gamma_1)$, for different values of the
positron energy $\gamma_+$, plasma density parameter
$\gamma^2_p$ is shown in figures 7 and 8.
Integrating over the solid angle $d\Omega_k$
the total cross section $\sigma_a$ for pair
annihilation can be determined.  The variation of $\sigma_a$
with $\gamma_+$ for different values of $\gamma^2_\rho$ is shown
in figure (9).  It is seen that the cross section for pair
annihilation increases with an increase in the plasma density.

\section{Pair Production}

The creation of an electron and a positron by the annihilation
of two photons is depicted in Figure (10).  Again using crossing
symmetry wherein one replaces $p_i \rightarrow - p_+, \ p_f
\rightarrow p_-, \ k \rightarrow k_1$ and $k^\prime \rightarrow -k_2$
in the Compton scattering matrix, the total cross section for the
process of pair production can be determined.
\begin{figure}
\plotone{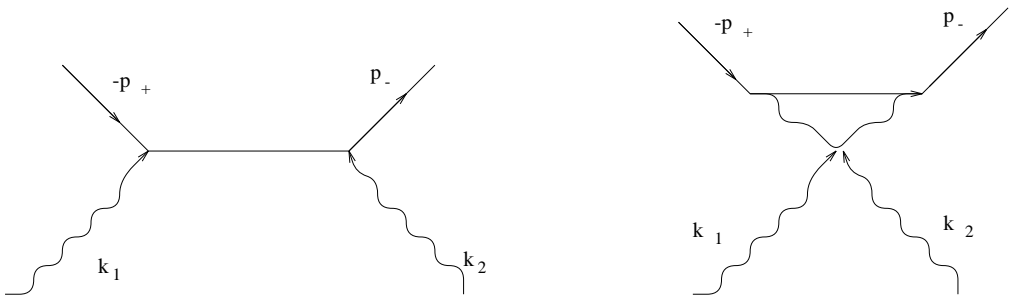}
\caption{Feynman diagrams for pair production.}
\end{figure}
 More directly, it is known that the total
 cross section $\sigma_p$ for pair production is related
to $\sigma_a $ (Reinhardt 1992, 1994) as: $\sigma_p = 2(\gamma^2 _+ - 1)
\sigma_a/\gamma^2_+$.  The variation of $\sigma_p$ with $\gamma_+$ for
different values of $\gamma^2_p$ is shown in figure (11).  One
can see the threshold behaviour of the pair production
cross section as well as its increase with an increase of the
plasma density.
\begin{figure}
\plottwo{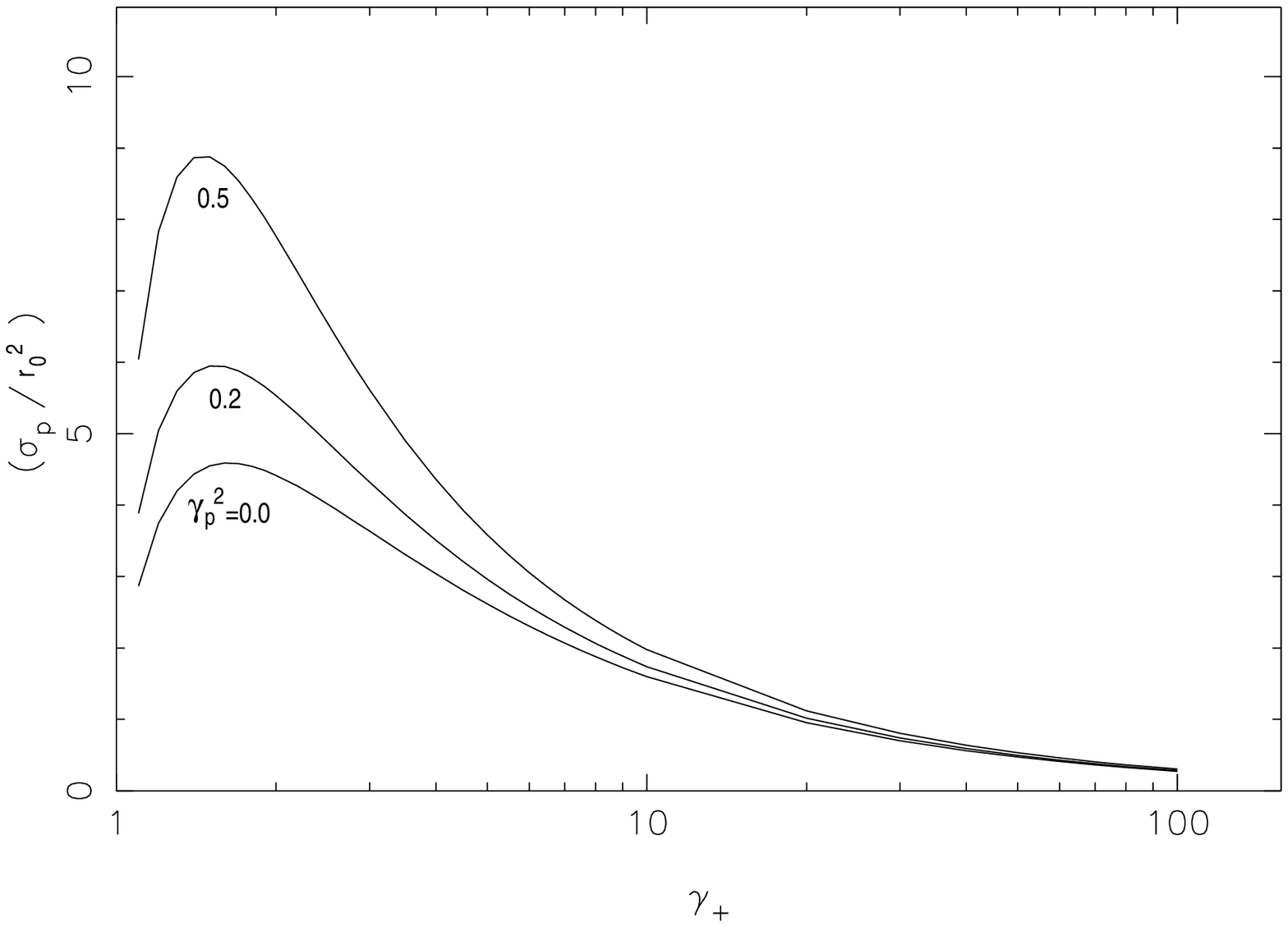}{fig12.ps}
\caption{Variation of the pair production cross section $\sigma_p$ with
positron energy $\gamma_+$ for different values of the plasma density 
parameters $\gamma_p^2.$}

\end{figure}

The enhancement of the pair annihilation and production
cross sections and the decrement of the Compton scattering
cross section (for small $\theta$) in a plasma are the consequences of the finite mass
that a photon acquires in a plasma.  Typical plasma densities at
which the effects are appreciable are of the order of $10^{30}$
cm$^{-3}$.  Such a density, for example, exists in the early
universe when its temperature was $\sim 10^{10} K$.  The highly
anisotropic nature of Compton scattering with reduced
cross sections for forward scattering in a plasma will have important consequences in
several astrophysical situations.

\begin{quote}
\verb"The author gratefully acknowledge discussions with"\\
\verb"Dr. R. T. Gangadhara, Dr. P. Ponda and Mr. B. A. Varghese."\\
\end{quote}

\section{References}

\begin{quote}
\verb"Harris,E.G., 1975 in Introduction to Modern Theoretical "\\
\verb"Physics, Vol.2, (John Wiley and Sons, New York), 594."\\

\verb"Krishan, V.,  1999 in Astrophysical Plasmas and Fluids "\\
\verb"(Kluwer Academic Publishers, Netherlands)."\\

\verb"Landau, L.D., Lifshitz, E.M., 1960 in Electrodynamics of"\\
\verb"continuous Media (translated by J.B.Sykes and J.S.Bell, "\\
\verb"Addison-Wesley, Reading, Mass., Section 61)."\\

\verb"Reinhardt, G., 1994 in Quantum Electrodynamics, second "\\
\verb"edition (Springer-Verlag Berlin Heidelberg 1992)."\\

\verb"Rose, W.K., 1973 in Astrophysics (Holt, Rinehart and Winston,"\\
\verb"Inc., New York)."\\

\verb"Wolfgang, B., Fabian, A.C., Giovannelli, F. 1990 i an"\\ 
\verb"Physical Processes in Hot Cosmic Plasmas (NATO ASI Series),"\\
\verb"305."\\
\end{quote}
\end{document}